\documentclass[%
preprint,
]{revtex4-2}

\usepackage{graphicx}
\usepackage{dcolumn}
\usepackage{bm}
\usepackage{natbib}

\begin{document}

\title{\textbf{Was Benoit Mandelbrot a hedgehog or a fox?} 
}%

\author{Rosario N. Mantegna}
\email{email: rosario.mantegna@unipa.it}
\affiliation{
Dipartimento di Fisica e Chimica Emilio Segrè, Università degli Studi di Palermo, Palermo, Italy}
\affiliation{Complexity Science Hub, Vienna, Austria}

\date{\today}

\begin{abstract}
Benoit Mandelbrot’s scientific legacy spans an extraordinary range of disciplines—from linguistics and fluid turbulence to cosmology and finance—suggesting the intellectual temperament of a “fox” in Isaiah Berlin’s famous dichotomy of thinkers. This essay argues, however, that Mandelbrot was, at heart, a “hedgehog”: a thinker unified by a single guiding principle. Across his diverse pursuits, the concept of scaling—manifested in self-similarity, power laws, fractals, and multifractals—served as the central idea that structured his work. By tracing the continuity of this scaling paradigm through his contributions to mathematics, physics, and economics, the paper reveals a coherent intellectual trajectory masked by apparent eclecticism. Mandelbrot’s enduring insight in the modeling of natural and social phenomena can be understood through the lens of the geometry and statistics of scale invariance.\\

L’h\'eritage scientifique de Beno\^it Mandelbrot couvre un spectre exceptionnel de disciplines — de la linguistique et de la turbulence des fluides \`a la cosmologie et \`a la finance — ce qui semble d’abord l’apparenter \`{a} la figure du “renard” dans la c\'{e}l\`{e}bre dichotomie des penseurs propos\'ee par Isaiah Berlin. Cet article soutient toutefois que Mandelbrot rel\`eve plus fondamentalement du type du “h\'erisson”, en tant que penseur structur\'e par un principe unificateur unique. Au fil de travaux en apparence tr\`es divers, le concept d’\'echelle — d\'eclin\'e \`a travers l’auto-similarit\'e, les lois de puissance, les fractales et les multifractales — constitue l’id\'ee centrale organisant son œuvre. En retra\c{o}ant la continuit\'e de ce paradigme de l’\'echelle dans ses contributions aux math\'ematiques, \`a la physique et \`a l’\'economie, l’article met au jour une trajectoire intellectuelle coh\'erente, masqu\'ee par un \'eclectisme apparent. L’apport durable de Mandelbrot \`a la mod\'elisation des ph\'enomènes naturels et sociaux peut ainsi \^etre interpr\'et\'e \`a l’aune de la g\'eom\'etrie et des statistiques de l’invariance d’\'echelle.
\end{abstract}


\maketitle


\section*{\label{sec:level1}Introduction}

Benoit Mandelbrot was a prominent figure in the scientific landscape of the twentieth century and the early years of the twenty-first. Throughout his career, he characterized himself as a scientific “maverick” \cite{mandelbrot2012fractalist}. A scientific maverick may be defined as a researcher who challenges established paradigms, frequently working across disciplinary boundaries or outside the academic mainstream, and who, in successful instances, reshapes the prevailing understanding within the scientific community. Over the course of his career, Mandelbrot made significant contributions to a diverse array of fields, ranging from linguistics to finance. His intellectual pursuits were, in fact, exceptionally wide-ranging and inherently multidisciplinary.

The present note undertakes an examination of Mandelbrot’s scientific legacy with the aim of determining whether his work is more appropriately interpreted as the expression of a generalist orientation—marked by the pursuit of diverse scientific problems—or as the manifestation of a specialist commitment to a unifying conceptual principle underlying his modeling practices. The analysis is situated within the interpretive framework provided by Isaiah Berlin’s essay The Hedgehog and the Fox \cite{berlin2013hedgehog}, which offers a useful lens through which to interrogate the tension between intellectual breadth and conceptual unity in Mandelbrot’s oeuvre.

\section{\label{sec:level2} The Hedgehog and the Fox}

In 1951, the same year that Mandelbrot published his first scientific article \cite{mandelbrot1951adaptation}, Isaiah Berlin, the Russian-British social and political theorist, philosopher, and historian of ideas, published The Hedgehog and the Fox \cite{berlin2013hedgehog}. The essay opens with a reflection on a fragment attributed to the Greek poet Archilochus: “The fox knows many things, but the hedgehog knows one big thing.” Berlin interprets this aphorism as articulating a profound distinction among thinkers, writers, and, more generally, human beings. As he explains:

{\it {`` But, taken figuratively, the words can be made to yield a sense in which they mark one of the deepest differences which divide writers and thinkers, and, it may be, human beings in general. For there exists a great chasm between those, on one side, who relate everything to a single central vision, one system, less or more coherent or articulate, in terms of which they understand, think and feel -- a single, universal, organising principle in terms of which alone all that they are and say has significance -- and, on the other side, those who pursue many ends, often unrelated and even contradictory, connected, if at all, only in some de facto way, for some psychological or physiological cause, related to no moral or aesthetic principle. These last lead lives, perform acts and entertain ideas that are centrifugal rather than centripetal; their thought is scattered or diffused, moving on many levels, seizing upon the essence of a vast variety of experiences and objects for what they are in themselves, without, consciously or unconsciously, seeking to fit them into, or exclude them from, any one unchanging, all-embracing, sometimes self-contradictory and incomplete, at times fanatical, unitary inner vision. The first kind of intellectual and artistic personality belongs to the hedgehogs, the second to the foxes; ..."}} \cite{berlin2013hedgehog}.

Berlin thus delineates two archetypal modes of intellectual engagement: the hedgehog, who interprets reality through the lens of a single, overarching vision, and the fox, who draws insight from a plurality of experiences and perspectives. This conceptual dichotomy provides a useful framework for evaluating the nature of Mandelbrot’s scientific enterprise and the extent to which his work reflects the attributes of either archetype.

For most writers and thinkers, classification along the hedgehog–fox spectrum (or some intermediate position within it) is relatively unproblematic. However, there are instances in which such a binary categorization proves inadequate. One such case, discussed extensively by Berlin himself, is that of Tolstoy. In Berlin’s words:

{\it {"But when we come to Count Lev Nikolaevich Tolstoy, and ask this of him -- ask whether he belongs to the first category or the second, whether he is a monist or a pluralist, whether his vision is of one or of many, whether he is of a single substance or compounded of heterogeneous elements --there is no clear or immediate answer"}} \cite{berlin2013hedgehog}. 

Analogously, a similar interpretive challenge arises when one critically examines the work of Benoit B. Mandelbrot. His intellectual legacy resists a straightforward classification: while his oeuvre displays the breadth and diversity characteristic of the fox, it simultaneously reveals a deep and persistent adherence to a unifying conceptual vision — an overarching role of the scaling concept — that aligns him with the hedgehog.

In fact, Mandelbrot’s research contributions span an extraordinary range of disciplines. He published significant works in information theory \cite{mandelbrot1951adaptation,mandelbrot1965leo}, linguistics \cite{mandelbrot1961theory}, hydrology \cite{mandelbrot1968noah}, stochastic processes \cite{mandelbrot1961stable,mandelbrot1968fractional,mandelbrot2003some}—with particular emphasis on long-memory phenomena—geometrical structures with non-integer dimensions \cite{mandelbrot1967long}, turbulence \cite{mandelbrot1974intermittent}, geophysics \cite{mandelbrot1969some}, astrophysics \cite{mandelbrot1975modele}, multifractals \cite{mandelbrot1974intermittent}, finance \cite{mandelbrot1963variation,mandelbrotFisherCalvet}, and economics \cite{mandelbrot1960pareto}. This extensive and diverse body of work—of which only the most prominent areas are cited here—could readily lead one to characterize Mandelbrot as a quintessential “fox,” a scientist who, in Berlin’s words, “performs acts and entertains ideas that are centrifugal rather than centripetal” \cite{berlin2013hedgehog}.

Yet, despite the manifestly interdisciplinary (or even multidisciplinary) nature of his scientific production, a coherent and persistent leitmotiv emerges across Mandelbrot’s career. Virtually all of his studies address systems, processes, and phenomena governed by some form of scaling behavior \cite{kadanoff1990scaling,stanley1999scaling}. This unifying focus on scaling —manifested in the exploration of self-similarity, fractal geometry, power-law distributions, up to more complex concepts such as multifractals — suggests that, beneath the apparent plurality of topics, Mandelbrot’s intellectual endeavor was driven by a single, organizing vision: the search for evidence highlighting a pervasive role of scaling in nature and in society.

\section{\label{sec:level3} Scaling in Mandebrot's works}

The concept of scaling is multifaceted \cite{kadanoff1990scaling}. In its most fundamental sense, it refers to the idea of scale invariance. When scale invariance is present, a system or phenomenon exhibits similar patterns when observed at different scales. The most striking manifestation of scale invariance is self-similarity—that is, the recurrence of certain geometrical or stochastic structures across multiple scales.

A paradigmatic example of scaling appears in random processes governed by power-law (or hyperbolic) probability density functions. For such distributions, rescaling the variable does not alter the form of the density function (apart from a multiplicative factor), implying that power-law distributions are inherently scale invariant 
\cite{newman2005power}.

One of Mandelbrot's first studies investigated Zipf’s law, an empirical relationship describing the frequency of words in literary texts, which is consistent with a power-law distribution \cite{mandelbrot1954structure,mandelbrot1961theory}.

His later discovery that the log-price changes of commodities traded in competitive markets follow probability distributions with markedly non-Gaussian tails \cite{mandelbrot1963variation} — again exhibiting power-law behavior — motivated him to “extend the notion of scaling to tackle several noise phenomena.”

Another milestone in Mandelbrot’s career was his work on the geometric properties of the coast of Britain, through which he introduced the concept of fractal objects and fractal dimensions \cite{mandelbrot1967long}. Geometric and stochastic fractals are self-similar structures that obey scaling relationships.

Mandelbrot illustrated geometric fractals by directing attention to mathematical sets and functions such as the Weierstrass curve, Koch curve, and Cantor dust. For stochastic fractals, he focused on random processes \cite{mandelbrot1983fractal}. In particular, he emphasized that the diffusive nature of Brownian motion implies a form of temporal scaling. In Mandelbrot’s view, a Brownian random walk is a process characterized by a well-defined fractal dimension.

He further demonstrated that the fractal nature of the Gaussian random walk was not unique. In fact, he expanded the family of stochastic fractals along two key directions. The first was the introduction of fractional Brownian motion, a stochastic process that allows for both superdiffusion and subdiffusion relative to standard diffusion, and that possesses a non-integer fractal dimension \cite{mandelbrot1968fractional}.

The second was the recognition and popularization of stable non-Gaussian random processes obeying a generalized version of the Central Limit Theorem. Mandelbrot was among the first scholars to suggest that real-world phenomena could often be more accurately described by stable non-Gaussian processes \cite{samorodnitsky1994stable}, that is, random processes governed by a generalized form of the central limit theorem. The existence of such generalized forms was established by Paul Lévy, Boris Vladimirovich Gnedenko, and Andrey Nikolaevich Kolmogorov \cite{gnedenko1954limit}. It is worth noting that Mandelbrot attended Lévy’s lectures while studying at the École Polytechnique, an experience that deeply influenced his later work \cite{mandelbrot1983fractal}.

Mandelbrot’s proposal to use stable non-Gaussian processes for modeling real-world phenomena required a genuine paradigm shift. Indeed, all stable non-Gaussian random variables are characterized by infinite variance, meaning that conventional statistical measures can become ill-defined or even meaningless. While this rendered the quantification of many aspects of such processes problematic, or seemingly impossible, Mandelbrot was undeterred. He analyzed these processes, both theoretically and empirically, through the lens of scaling, demonstrating that their scaling properties remain perfectly well-defined even when traditional statistical measures cannot be used.

Mandelbrot highlighted the existence of a more sophisticated form of scaling even in processes that appear to lack self-similarity, namely, multifractal processes \cite{mandelbrot1974intermittent,Parisi1985singularity,benzi1984multifractal,halsey1986fractal}. In a multifractal process, the complexity of the scaling structure is expressed through a continuous spectrum of exponents, each associated with a particular fractal dimension. Mandelbrot noted that the generalization from fractal sets to multifractal measures “involves the passage from geometric objects that are characterized primarily by one number—namely, a fractal dimension—to geometric objects that are characterized primarily by a function. \cite{mandelbrot1989multifractal}” Despite this shift, both fractals and multifractals retain a form of self-similarity. In fractals, self-similarity pertains to the geometric set itself, whereas in multifractals, it applies to the measure (in the mathematical sense) that characterizes the system under investigation. Mandelbrot developed the first example of a multifractal when studying fully developed turbulence, a phenomenon that cannot be described by a single self-similar deterministic or stochastic process \cite{mandelbrot1974intermittent}. He later extended the multifractal framework to model financial asset returns in competitive markets \cite{mandelbrotFisherCalvet}.

Regarding the modeling of return dynamics in such markets, it is worth noting that Mandelbrot’s views evolved over time—from his 1963 model based on stable non-Gaussian distributions \cite{mandelbrot1963variation} to his 1997 model involving a multifractal process \cite{mandelbrotFisherCalvet}. The two approaches differ crucially in how they treat the second moment of the return probability density function. In the stable non-Gaussian model, the variance is infinite, whereas in the multifractal model the best agreement with empirical data is obtained when the variance is finite. Thus, Mandelbrot revised his stance on the finiteness of volatility in asset returns but maintained his belief in an underlying self-similar regularity. In the stable non-Gaussian model, self-similarity is present in the dynamics of asset returns themselves, while in the multifractal model, it governs the scaling behavior across different time horizons, leading to a progressive convergence toward a Gaussian process.

Mandelbrot himself summarized the different models of return dynamics he proposed over the years in books (see, for example, ref. \cite{mandelbrot1997fractals}) and papers (in our opinion the most representative reference for this aspect is \cite{mandelbrot2001scaling}). 

In ref. \cite{mandelbrot2001scaling} Mandelbrot presents his models within a framework where price variation is a self-affine random process with events occurring at time instants described by a multifractal process. He addressed this family of stochastic processes as Brownian motion multifractal time BMMT. In particular, the paper introduces simplified recursive constructions—termed cartoon Brownian motions—designed to isolate and visualize the essential scaling properties observed in financial data.

The paper reiterates the empirical inadequacy of classical Brownian motion, which assumes Gaussian returns, constant volatility, and independence. It discusses earlier models proposed by Mandelbrot, namely Lévy stable processes (capturing heavy tails) \cite{mandelbrot1963variation} and fractional Brownian motion (capturing long-range dependence) \cite{mandelbrot1965classe,mandelbrot1968fractional}, acknowledging that these earlier models address only subsets of the empirical features characterising asset return stylised facts. A key characteristic of the paper is that all previous models are embedded in a unified framework highlighting time multifractality of events.

Specifically, he presents a number of random generations constructed through recursive interpolation on an increasingly refined time grid, using a fixed three-interval generator with asymmetric rescaling of time and price. Although grid-bound and artificial, these constructions are self-affine and exhibit a wide range of behaviours depending on generator parameters. Mandelbrot introduces a summary (named by him a ``phase diagram") that relates these behaviors to distinct stochastic processes. He considers, by using a rather personal terminology, so-called Fickian (Brownian), unifractal (fractional Brownian), mesofractal (discontinuous, Lévy-like), and multifractal random processes \cite{mandelbrot2001scaling}. In his view all these models are random processes that can be seen as special cases of his generically defined BMMT. In the discussion of these different specifications of the general model he acknowledges that in the description of empirically observed regimes of asset price changes the most appropriate model needs to reproduce the combination of volatility clustering, intermittency, and sharp price spikes characteristic of financial markets.

In summary, Mandelbrot revised a few times his view on the best model describing asset returns but all models he proposed included a form of scaling. Scaling was always an essential ingredient of models he proposed \cite{mandelbrot1997fractals}.

\section{\label{sec:level5} Conclusions}
Was Mandelbrot a hedgehog or a fox?
Mandelbrot’s own answer to this question was that he was a hedgehog. In the foreword to his masterpiece The Fractal Geometry of Nature \cite{mandelbrot1983fractal}, he wrote:

“My first scientific publication came out on April 30, 1951. Over the years, it had seemed to many that each of my investigations was aimed in a different direction. But this apparent disorder was misleading: it hid a strong unity of purpose, which the present Essay, like its two predecessors, is intended to reveal.”

The revelation to which Mandelbrot referred concerned the role of scaling in both mathematical and real-world systems and processes. Scaling was the leitmotif he perceived as underlying a vast range of natural and artificial structures—so fundamental, in fact, that he regarded it as one of the foundational principles of nature itself, as the title of his book suggests.

Throughout his career, Mandelbrot conducted research and developed models that challenged the prevailing view in the scientific community that scaling and self-similarity were merely rare curiosities—phenomena to be confined, as he put it, to a “Mathematical Art Museum” or, worse, a “Gallery of Monsters \cite{mandelbrot1983fractal}”. His position was the very opposite: for him, scaling and self-similarity were essential concepts for understanding the fundamental structure and behavior of both mathematical objects and real-world systems.

By treating scaling and self-similarity as indispensable principles of nature, Mandelbrot embodied the hedgehog’s single-minded focus—seeking, across multiple scientific disciplines, empirical evidence and theoretical models demonstrating that many systems (indeed, in his view, most of the significant ones) are best described through scale-invariant frameworks rather than by simplistic, reductionist approaches that overlook the intrinsic self-similarity of processes and structures.

Through his persistent and authoritative contributions, Mandelbrot’s lifelong effort to illuminate the nature and importance of scaling and self-similarity has profoundly influenced modern science. Today, the study of scale-invariant processes across disciplines bears the unmistakable imprint of his vision.


\end{document}